\documentstyle[11pt,newpasp,twoside]{article}
\begin{document}

\title{Stellar Populations, Bars and Secular Evolution in Late--Type Galaxies}
\author{Dimitri A. Gadotti and Sandra dos Anjos}
\affil{Departamento de Astronomia, Universidade de S\~ao Paulo, Av. Miguel
       Stefano, 4200, S\~ao Paulo -- SP, Brasil, CEP 04301-904}

\begin{abstract}
We have done a robust statistical analysis of UBV color profiles 
of 257 Sbc barred and unbarred galaxies. We found that 
there is an excess of barred galaxies among the objects with null
or positive (bluish inward) color gradients, which seems to indicate that 
bars act as a mechanism of homogenization of the stellar population along 
galaxies. Moreover, the relationship found between total and bulge colors shows
that, in the process of homogenization, the stellar population of bulges are 
getting bluer, whereas the total color of galaxies remains the same.
These characteristics are expected in a secular evolutionary 
scenario, and seem incompatible with both the monolithic and the 
hierarchical scenarios for spiral galaxy formation.
\end{abstract}

\section{Color Gradients in Late--Type Barred Galaxies}
We estimated $(B-V)$ and $(U-B)$ color gradients, as well as total and bulge 
mean colors, for each galaxy of a sample of 257 barred and unbarred Sbc 
systems, using the catalog of Longo \& A. de Vaucouleurs (1983,1985), 
which presents photoelectric aperture photometry data. The gradients were 
calculated through robust statistical methods, which consist essentially
in applying the Least Median of Squares regression (Rousseeuw 1984). 
Moreover, we have done a comparative photometric 
study through CCD images to atest the validity of the results obtained.

Our results, considering only face--on galaxies, 
indicate that $\sim$ 59\% of late--type galaxies have negative 
color gradients, $\sim$ 27\% have no gradients, and $\sim$ 14\%  have positive 
gradients. Furthermore, there is an excess of barred galaxies
among systems with zero or positive color gradients (Table 1). Thus, it seems 
that bars act as a mechanism of homogenization of the stellar population 
along late--type galaxies, a result which puts in trouble the monolithic 
scenario of Eggen, Lynden--Bell, \& Sandage (1962).

\begin{table}
\caption{Distribution of the face--on galaxies in our sample in relation to the three regimes of gradients, with the
median of their bulge and total mean colors.}
\begin{center}
\begin{tabular}{lllllll}
\tableline
Gradient & Color   &   N   & Sample & Barred & \hskip0.5cm Bulge & \hskip0.5cm Total \\
	 &	   &  \hskip-0.1cm (1) &  \hskip0.4cm (2)   &  \hskip0.4cm (3)  & & \\
\tableline
Negative & $(B-V)$ &  \hskip-0.1cm 78  &  \hskip0.3cm 63\%  &  \hskip0.3cm 75\% &  0.74 $\pm$ 0.01 &  0.53 $\pm$ 0.01 \\
Null	 &	   &  \hskip-0.1cm 32  &  \hskip0.3cm 26\%  &  \hskip0.3cm 91\% &  0.57 $\pm$ 0.02 &  0.57 $\pm$ 0.02 \\
Positive &	   &  \hskip-0.1cm 14  &  \hskip0.3cm 11\%  &  \hskip0.3cm 72\% &  0.36 $\pm$ 0.03 &  0.55 $\pm$ 0.05 \\
\tableline
Negative & $(U-B)$ &  \hskip-0.1cm 55  &  \hskip0.3cm 53\%  &  \hskip0.3cm 73\% &  0.24 $\pm$ 0.03 & \hskip-0.1cm -0.05 $\pm$ 0.02 \\
Null	 &	   &  \hskip-0.1cm 30  &  \hskip0.3cm 29\%  &  \hskip0.3cm 83\% & \hskip-0.1cm -0.05 $\pm$ 0.03 & \hskip-0.1cm -0.05 $\pm$ 0.03 \\
Positive &	   &  \hskip-0.1cm 19  &  \hskip0.3cm 18\%  &  \hskip0.3cm 90\% & \hskip-0.1cm -0.31 $\pm$ 0.07 &  0.05 $\pm$ 0.07 \\
\tableline
\tableline
\end{tabular}
\end{center}
{\it Note. --- }(1): number of galaxies in each regime; (2): fraction of the sample in each regime; (3): fraction of 
barred galaxies in each regime.
\end{table}

\section{Blue Bulges}
We found a correlation between total and bulge colors, which is a 
consequence of an underlying correlation between the colors of bulges 
and disks found by other authors (Peletier \& Balcells 1996). 
Moreover, one can see in Table 1 that the mean total color is 
approximately the same irrespective of the gradient regime,
while bulges are systematically bluer in galaxies with null or positive 
gradients. These results seem to indicate that bars act 
changing the color of bulges due to secular dynamical processes (see, 
e.g., Friedli \& Martinet 1993), 
turning these bulges bluer, while disks are 
getting redder as a consequence of passive evolution.
These effects are expected in the secular evolutionary scenario, 
but they are not consistent with the monolithic and hierarchical scenarios
(e.g., Kauffmann, Guiderdoni, \& White 1994) for spiral galaxy formation.

\section{Further Results}
We also found no correlation between color and 
chemical abundance gradients, suggesting that the color 
gradients are more sensitive to the age 
rather than to the metallicity of the stellar population. Further
results and details will appear in a forthcoming paper
(Gadotti \& dos Anjos, in preparation).

\acknowledgments
We acknowledge the Conselho Nacional de Pesquisa e Desenvolvimento (CNPq), 
the NExGal -- ProNEx, the Funda\c c\~ao de Amparo \`a Pesquisa do 
Estado de S\~ao Paulo (FAPESP; grant 99/07492-7), 
and NSF grant AST-9900789 
for the financial support.

\end{document}